\documentclass[10pt,aps,pre,showpacs,twocolumn]{revtex4-1}

\usepackage{graphicx}

\usepackage{amsmath}

\begin{document}



\title{Instability of Planetary Flows using Riemann Curvature:
a Numerical Study
}


\author{Richard Blender}
\affiliation{Meteorologisches Institut,  
Universit\"at Hamburg, Hamburg, Germany}

\date{\today}

\begin{abstract}
The instability of ideal non-divergent zonal flows 
on the sphere is determined numerically by the
instability criterion of Arnol'd (1966) for the  sectional curvature.
Zonal flows are unstable for all perturbations
besides for a small set which are in approximate resonance.
The sectional curvature scales with $m/\ell$ 
for large total and zonal wave numbers  $\ell$ and $m$
of the perturbations. 
The planetary rotation is stable 
and the presence of rotation reduces the instability of perturbations.
\end{abstract}



\maketitle

\section{Introduction}


Arnol'd formulated ideal
two-dimensional hydrodynamics 
as a geodesic in state space \cite{Arnold:1966,Arnold:1978}.
The dynamics is given by a volume reserving diffeomorphism on a torus
(double periodic domain). 
The torus becomes a Riemannian manifold when the metric given by 
the kinetic energy is added.
The instability of flows can be assessed by the Jacobi equation and
a flow is considered to be unstable if the sectional curvature 
in the Jacobi equation is negative.
Arnol'd found that the flow on a torus is unstable
and used this  to explain the limited predictability of 
atmospheric flows. 

%
%

%
%
Spherical flows have been considered in
\cite{Lukatskii:1979,Lukatskii:1988, Arakelyan:1989, 
Dowker:1990, Yoshida:1997}. 
Arakelyan and Savvidy \cite{Arakelyan:1989} 
and Dowker and Mo-Zheng \cite{Dowker:1990}
derive the sectional curvature in the spherical harmonics basis. 
Yoshida \cite{Yoshida:1997} approached the problem
by a transformation of the 
spherical harmonics.
Kambe \cite{Kambe:1998} reviews the method 
and mentions some hydrodynamic examples. 
Casetti et al. \cite{Casetti:2000} review the applications 
with an emphasis on statistical mechanics.


The aim of the present work is to derive numerical results
for the instability of nondivergent zonal flows
on the sphere.
The calculations are based on \cite{Arakelyan:1989} 
and  \cite{Dowker:1990}.
The results are visualized for zonal flows with low total wave 
numbers which correspond to observations. 
The planetary rotation is considered by 
adding a solid body rotation represented by
the stream-function $Y_{10}$
to the zonal flow stream-functions. 
This paper is organized as follows:
In Section \ref{Sec_Seccurv} the 
instability of spherical flows is described in terms
of the sectional curvature which is based on 
the Riemann curvature tensor.
In Section \ref{Results} the results are presented 
and in Section \ref{Summary} the findings are summarized and 
discussed.

\section{Sectional curvature } \label{Sec_Seccurv}

This section summarizes 
the mathematical properties of non-divergent flows on
the sphere and  introduces the structure constants
to derive the Riemann curvature tensor and
the sectional curvature.
%
%
Explicit expressions for the sectional curvature on the sphere
are given in \cite{Arakelyan:1989}
and \cite{Dowker:1990}.
Here we follow \cite{Arakelyan:1989}. 
The  Riemann tensor
and the sectional curvature derived in \cite{Dowker:1990} 
can be found in  the Appendix.
We consider non-divergent flows on the sphere $S^2$
with unit radius,
which are denoted by SDiff$S^2$,
the volume preserving diffeomorphisms on the sphere.
The coordinates are the latitude 
$\sigma_1=-\cos \theta$, and
the longitude $\sigma_2=\lambda$; 
the North Pole is at $\theta=0$. 


For the stream-functions $\psi$ and $\eta$
divergence-free 
vector fields ${\bf u}, {\bf v}$ are determined by
\begin{equation}
        u^\alpha= \varepsilon^{\beta \alpha} \partial_\beta \psi,
        \qquad
        v^\alpha= \varepsilon^{\beta \alpha} \partial_\beta \eta,
\end{equation}
for $\alpha,\beta=1,2$ and the anti-symmetric symbol 
$\varepsilon^{\beta \alpha}=1$ for 
$(\beta, \alpha)=(1,2)$, $-1$ for $(2,1)$
and  0 otherwise. The derivative $\partial_\alpha$ is with respect to 
the coordinate $\sigma_\alpha$.


A Riemannian manifold is a differentiable manifold 
with a metric $\langle {\bf u}, {\bf v} \rangle$ 
for vector fields ${\bf u}$ and ${\bf v}$.
This is chosen proportional to the kinetic energy, 
\begin{align}
        \langle {\bf u},{\bf v} \rangle 
       & = \int \nabla \psi \cdot \nabla \eta \, \mbox{d}A
        \nonumber
        \\
        & = - \int  \psi \nabla^2 \eta \, \mbox{d}A
        = (\psi, \eta)
        \label{KEmetric}
\end{align}
where the last term defines the metric $(,)$ for the stream-functions. 
The area element  is
$ \mbox{d}A = \mbox{d} \sigma_1
\mbox{d} \sigma_2
=
\sin \theta \,  \mbox{d} \theta \,  \mbox{d} \lambda $. 


An orthogonal basis for the stream-functions on the sphere
is given by the spherical harmonics
\begin{equation}
        Y_{\ell m} 
        = C_{\ell m}
        e^{i m \lambda}
        P_{\ell}^{|m|}(\cos \theta)
\end{equation}
\begin{equation}
        C_{\ell m}
        = (-1)^m
        \sqrt{
        \frac{2\ell+1}{4\pi}
        \frac{(\ell-|m|)!}{(\ell+|m|)!}
        }
\end{equation}
with the Condon and Shortley phases $(-1)^m$ not included
in the associated Legendre polynomials $P_{\ell}^{|m|}(\cos \theta)$.
The spherical harmonics are orthogonal
\begin{equation}
        (Y_{\ell_1 m_1}, Y_{\ell_2 m_2})
        = - (-1)^{m_1} \ell_1(\ell_1+1)
        \delta_{\ell_1,\ell_2}
        \delta_{m_1,-m_2}
\end{equation}
in the kinetic energy metric (\ref{KEmetric}).


A Poisson algebra can be defined
for stream-functions $\psi$ and $\eta$
\begin{equation}        \label{PApsieta}
        [ \psi, \eta ]_P
        = \partial_1 \psi \partial_2 \eta - \partial_2\psi \partial_1 \eta 
\end{equation}
with the structure constants $G^{\ell_3 m_3}_{\ell_1 m_1 \ell_2 m_2}$
\begin{equation}         \label{scpa}
        [ Y_{\ell_1 m_1}, Y_{\ell_2 m_2} ]_P
        = G^{\ell_3 m_3}_{\ell_1 m_1 \ell_2 m_2}
        Y_{\ell_3 m_3}
\end{equation}
%
%
Since the structure constants on the sphere are imaginary, 
it is convenient to use
the imaginary parts (the so-called real structure constants) $g$
\begin{equation}                \label{scla}
        G^{\ell_3 m_3}_{\ell_1 m_1 \ell_2 m_2}  
        = -i (-1)^{m_3}
         g^{\ell_3 -m_3}_{\ell_1 m_1 \ell_2 m_2}
\end{equation}
Arakelyan and Savvidy \cite {Arakelyan:1989} 
derive  the real structure constants $g$
\begin{widetext}
\begin{multline} 
\label{gAS}
        \left(
        \frac{4\pi}{(2 \ell_1+1)(2\ell_2+1)(2\ell_3+1)}
        \right)^{1/2}
        g^{\ell_3 m_3}_{\ell_1 m_1 \ell_2 m_2}
        =
        \\
        = m_2
        \sum_{k_1}
        [2(\ell_1-2k_1-1)+1]
        \sqrt{\frac{(\ell_1-|m_1|) 
         \dots (\ell_1-|m_1|-2k_1)}
               {(\ell_1+|m_1|) 
               \dots (\ell_1+|m_1|-2k_1)}}
        \left(
        \begin{array}{ccc}
         \ell_1-2k_1-1   &   \ell_2   &   \ell_3 \\
        m_1 & m_2 & m_3 
        \end{array}
        \right)
        \left(
        \begin{array}{ccc}
         \ell_1-2k_1-1   &   \ell_2   &   \ell_3 \\
        0 & 0 & 0 
        \end{array}
        \right)
        \\
        -m_1
        \sum_{k_2}
        [2(\ell_2-2k_2-1)+1]
        \sqrt{\frac{(\ell_2-|m_2|) 
        \dots (\ell_2-|m_2|-2k_2)}
               {(\ell_2+|m_2|) 
               \dots (\ell_2+|m_2|-2k_2)}}
        \left(
        \begin{array}{ccc}
         \ell_1 & \ell_2-2k_2-1    &   \ell_3 \\
        m_1 & m_2 & m_3 
        \end{array}
        \right)
        \left(
        \begin{array}{ccc}
         \ell_1  &  \ell_2-2k_2-1   &   \ell_3 \\
        0 & 0 & 0 
        \end{array}
        \right)
\end{multline}
\end{widetext}
with the 3-j symbols $(:::)$.


The instability of a flow $\psi$ 
with respect to perturbations $\eta$
is determined by the sectional curvature $K(\psi,\eta)$
in the Jacobi equation
\cite{Arnold:1978}.
In the spherical harmonics basis the two flows are 
represented as
\begin{equation}
        \psi= \sum_{\ell m} \psi^{\ell m} Y_{\ell m}, \quad
        \eta= \sum_{\ell m} \eta^{\ell m} Y_{\ell m}
\end{equation}
%
The flows  $\psi$ and $\eta$ are normalized, 
$\langle  \psi,\psi \rangle =1$, $\langle  \eta,\eta \rangle =1$, 
and orthogonal, 
$\langle  \psi,\eta \rangle =0$.

The sectional curvature in the basis of the spherical harmonics
is \cite{Dowker:1990}
\begin{equation} \label{KSC}
        K(\psi,\eta) = 
        -\sum
        R_{\ell_1 m_1 \ell_2 m_2 \ell_3 m_3 \ell_4 m_4}
       \psi^{\ell_1 m_1} 
       \eta^{\ell_2 m_2} 
       \psi^{\ell_3 m_3} 
       \eta^{\ell_4 m_4} 
\end{equation}
where the sum is over all wave numbers $\ell_i$ and $m_i$, 
$i=1,\dots,4$.


In the spherical harmonics basis, 
the Riemann tensor is determined by the structure constants 
\cite{Arakelyan:1989}
\begin{align}
          R_{\ell_1 m_1 \ell_2 m_2 \ell_3 m_3 \ell_4 m_4}
        &= \sum_{\ell m}
        (-1)^m \ell(\ell+1)
         \nonumber
        \\
        &[
                d^{\ell}_{\ell_1 \ell_3}
                d^{\ell}_{\ell_2 \ell_4}
                 G^{\ell m}_{\ell_1 m_1 \ell_3 m_3}
                G^{\ell -m}_{\ell_2 m_2 \ell_4 m_4}
        \nonumber
        \\                
               &-
                d^{\ell}_{\ell_1 \ell_4}
                d^{\ell}_{\ell_2 \ell_3}
                G^{\ell m}_{\ell_1 m_1 \ell_4 m_4}
                G^{\ell -m}_{\ell_2 m_2 \ell_3 m_3}
        \nonumber
        \\
               &+
                k^{\ell}_{\ell_3 \ell_4}
                G^{\ell m}_{\ell_1 m_1 \ell_2 m_2}
                G^{\ell -m}_{\ell_3 m_3 \ell_4 m_4}
        ]
\end{align}
with
\begin{equation}
        d^{\ell}_{\ell_1 \ell_2}
        =
        \frac{1}{2}
        \frac{\ell(\ell+1)
        -\ell_1(\ell_1+1)+\ell_2(\ell_2+1)}{\ell(\ell+1)}
\end{equation}
and
\begin{equation}
        k^{\ell}_{\ell_1 \ell_2}
        =
        \frac{1}{2}
        \frac{\ell(\ell+1)
        -\ell_1(\ell_1+1)-\ell_2(\ell_2+1)}{\ell(\ell+1)}
\end{equation}

%
%
%

In the Appendix the sectional curvature and
the Riemann tensor of \cite{Dowker:1990} are quoted.

\begin{figure}[h]
\includegraphics[width=8cm,
angle=0]{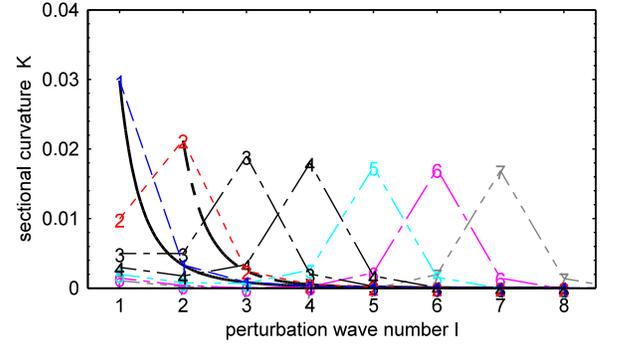}  
\caption{\label{fig1} 
Sectional curvature $K(\psi,\eta)$ for $\psi=Y_{L0}$ with $L$ as indicated
and perturbations $\eta_{\ell m}$
with $m=1$. 
Results from  \cite{Dowker:1990} 
are included
for $L=1$,  Eq. (\ref{K38}), solid-bold, 
and $L=2$, Eq. (\ref{K4038}) 
for  $\ell \geq2$,  dashed-bold.
}
\end{figure}

\section{Results}  \label{Results}


The numerical algorithm is first 
applied to the instability of the zonal flows $Y_{L0}$
for the reduced set of perturbations $\eta_{\ell m}$
with $m=1$ (Fig. \ref{fig1}). 
The analysis confirms previous results  
for $L=1$ and $L=2$ ($\ell \geq2$) \cite{Dowker:1990}.
For example for $L=1$ 
\begin{equation} \label{K38}
	K=\frac{3}{8\pi}\frac{1}{\ell^2 (\ell+1)^2 }
\end{equation}
and for $L=2$ 
\begin{equation}  \label{K4038}
   K=\frac{403}{8 \pi} \frac{1}{\ell^2 (\ell+1)^2 (2\ell+3) (2\ell-1) },
        \quad \ell \geq 2
\end{equation}
A quasi regular pattern emerges
where the stability is highest for perturbations with
total wave numbers close to the mean flow, $\ell \approx L$.

\begin{figure}[t]
\includegraphics[width=10cm,
angle=-90]{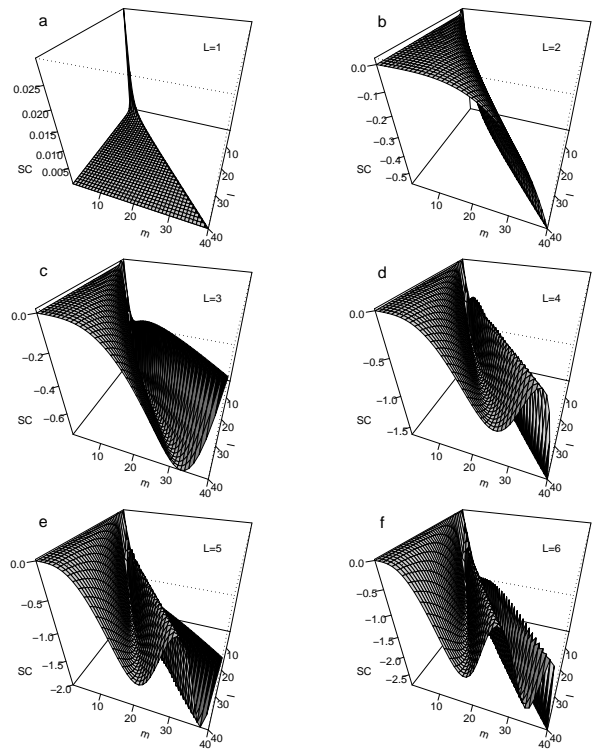}  
\caption{\label{fig2} 
Sectional curvature $K(\psi,\eta)$ for zonal flows $\psi=Y_{L0}$ 
and perturbations $\eta_{\ell m}$. 
(a) $L=1$, solid body rotation,
which is stable for all perturbation wave numbers $(\ell, m)$. 
(b)-(f) Sectional curvature for 
increasing total wave numbers $L=2,\dots,6$ as indicated.
}
\end{figure}

The stability of zonal flows $\psi=Y_{L0}$
with respect to arbitrary perturbations $\eta=Y_{\ell m}$ 
is presented in Fig. \ref{fig2}.
Note that the results for $m=1$ 
in Fig. \ref{fig1} are visible along the boundary
for low $L$.
For higher $L$ this becomes invisible since the instabilities 
increase and the scale changes accordingly.

Several findings in Fig. \ref{fig2} are noted:
\begin{itemize}
\item
The sectional curvatures scales with the argument $m/\ell$
for large wave numbers $\ell, m$.
Furthermore, the curvature varies weakly along these lines
(see for example $L=3$).
This property could be useful for 
expansions and approximations.
\item
The instability increases with $L$.
This supports the dominance of
low meridional wave numbers of zonal flows on planets.
\item
The stability of flows with even and odd total wave numbers
differs for the highest $m$:
for odd $L$ the curvature is less negative for high $m$,
whereas for even $L$ the instability is most negative for $m=\ell$.
Thus  flows with odd total wave numbers $L$ (symmetric
with respect to the equator) are less unstable.
\end{itemize}

\begin{figure}[h]
\includegraphics[width=7cm,
angle=0]{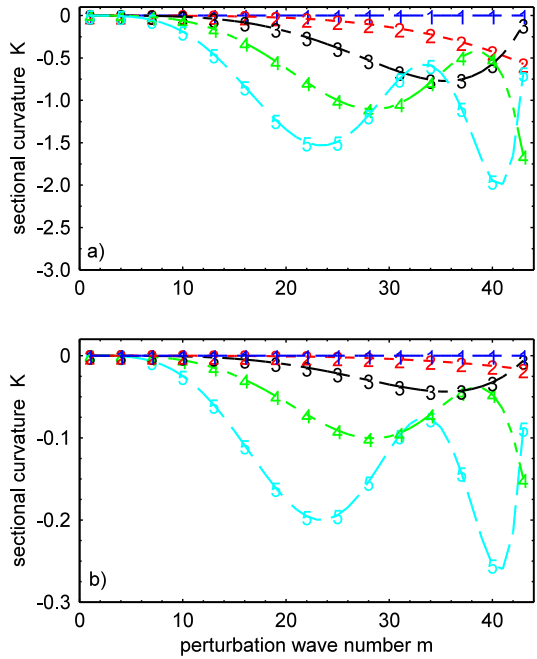}  
\caption{\label{fig3} 
Impact of rotation: 
a) Sectional curvature $K(\psi,\eta)$ for zonal flows 
$\psi=Y_{L0}$, $L=1,\dots,5$,
versus the zonal wave number $m$
of the perturbations $\eta^{\ell m}$ with constant $\ell=43$,
b) with solid body rotation superimposed to weak zonal flows,
$\psi=Y_{10} + 0.1 Y_{L0}$. 
}
\end{figure}

The above results show that the zonal flow $Y_{10}$
which corresponds to a solid body rotation is stable
with respect to all perturbations.
Thus planetary rotation might explain the observed stability  
of zonal slows.
Here we  consider the superposition of a rotation $Y_{10}$ 
with weak zonal flows (one order of magnitude smaller). 
The instabilities in Fig. \ref{fig3} are reduced by an order
of magnitude. Note that the reduction is not a simple rescaling
with one magnitude; it depends on $L$ and 
flows with small $L$ benefit 
more from the stabilization effect than those with higher $L$.

\section{Summary and Discussion}   \label{Summary}

The instability of ideal non-divergent spherical flows 
is determined numerically by the instability criterion of
Arnol'd \cite{Arnold:1966,Arnold:1978} for the  sectional curvature.
Using structure constants and the Riemann curvature tensor
in a  spherical harmonics basis the 
sectional curvature is determined numerically
for stationary zonal flows $Y_{L0}$ 
with meridional wave numbers $L$.
Zonal flows for $L \geq 2$ are unstable for all perturbations
besides for a small set with $m=1$.
For perturbations with large total and zonal wave numbers  
$\ell$ and $m$ 
the sectional curvature scales with $m/\ell$.

The planetary (solid body)  rotation given by $Y_{10}$  is stable. 
A superposition of intense planetary rotation reduces the
instability of zonal flows $Y_{L0}$ for $L>1$.
This result confirms the $\beta$-effect in geophysical fluid dynamics 
where $\beta$ is the meridional derivative of
the Coriolis parameter $f$, $\beta = df/d\phi$,
with latitude $\phi$ ($\phi=0$ at the equator).
The $\beta$-effect sustains Rossby waves 
and stabilizes geophysical turbulence.
Nevertheless, the stabilization is too weak
to compensate instabilities even
if the rotation is one order of magnitude more intense 
than the zonal flows.

This study cannot explain the observed
dominance of zonal flows on planets. 
There are two main reasons:
A first is that the assumption of a 
non-divergent ideal flow is too simplistic.
Realistic flow patters need to be simulated with
complex geophysical models (see for example \cite{Lian-2010}).
A second reason might be that the interpretation
of the sectional curvature as an indicator for instability is not yet fully understood (chaos can, for example, be induced by fluctuations
of positive curvatures as well \cite{Casetti:2000}).
%

\begin{acknowledgments}
These results have been presented at the EGU 2011, Vienna,
and during the Workshop on
Instabilities and Fluctuations of Geophysical Flows
in Hamburg, June 4-6, 2014, 
funded by the Deutsche Forschungsgemeinschaft.
\end{acknowledgments}




\appendix

\section{Riemann tensor by Dowker and Mo-zheng (1990) }

Dowker and Mo-zheng \cite{Dowker:1990} determine an
alternative 
closed expression for the negative imaginary part of the
structure constants   (\ref{scla}). 
These results are quoted (with a small correction)
since they turned out as numerically efficient.
\begin{equation}  \label{DMGC}            
        G^{\ell_3 m_3}_{\ell_1 m_1 \ell_2 m_2}  
        =-iC^{ \ell_1 \ell_2 \ell_3}_{ m_1 m_2 m_3 }
\end{equation}
%
%
\begin{align} \label{DMnew}
        C^{ \ell_1 \ell_2 \ell_3}_{ m_1 m_2 m_3 }
        =
        \frac{(-1)^{m_3}}{\sqrt{4\pi}}
        L_{123}
        \left(
        \begin{array}{ccc}
         \ell_1 & \ell_2 &   \ell_3 \\
        m_1 & m_2 & -m_3 
        \end{array}
        \right)
        \nonumber
        \\
        \times
        \left(
        \begin{array}{ccc}
         \ell_1 & \ell_2    &   \ell_3 \\
          1 &  -1 & 0 
        \end{array}
        \right)
\end{align}
with
\begin{equation}
        L_{123} 
        =
        \left[
        (2\ell_1+1)
        (2\ell_2+1)
        (2\ell_3+1)
        \ell_1(\ell_1+1)
        \ell_2(\ell_2+1)
        \right]^{1/2}
\end{equation}
The Riemann curvature components are in the notation of
\cite{Dowker:1990} 
\begin{align}
        R^{\ell_1 \ell_2 \ell_3 \ell_4}_{m_1 m_2 m_3 m_4}
        & = 
        \sum_{\ell m}
        (-1)^{m}
        \ell(\ell+1)
        \nonumber
        \\
        [      & -
                d^{\ell}_{\ell_1 \ell_3}
                d^{\ell}_{\ell_2 \ell_4}
                C^{\ell_1 \ell_3 \ell}_{m_1 m_3 m}
                C^{\ell_2 \ell_4  \ell}_{m_2 m_4 -m}
        \nonumber
        \\
              &+
                d^{\ell}_{\ell_1 \ell_4}
                d^{\ell}_{\ell_2 \ell_3}
                C^{\ell_1 \ell_4 \ell}_{m_1 m_4 m}              
                C^{\ell_2 \ell_3 \ell}_{m_2 m_3 -m}      
                \nonumber
              \\
              & -
                k^{\ell}_{\ell_3 \ell_4}
                C^{\ell_1 \ell_2 \ell}_{m_1 m_2 m}              
                C^{\ell_3 \ell_4 \ell}_{m_3 m_4 -m}             
        ]
\end{align}
Note that \cite{Dowker:1990} mention a different
expression (possibly a typo); the relationships 
(\ref{DMGC},\ref{DMnew}) were
verified numerically with (\ref{scla}) and (\ref{gAS}).


\bibliographystyle{unsrt}

\bibliography{Riem}


\end{document}